\documentclass[10.5pt,prx,aps,notitlepage,superscriptaddress,onecolumn]{revtex4-1}

\usepackage[utf8]{inputenc}
\usepackage[T1]{fontenc}
\usepackage[english]{babel}  

\usepackage{colortbl}
\usepackage{tabu,diagbox}

\usepackage{times}
\usepackage{dsfont}

\usepackage{amssymb,amsmath,amsfonts,amsthm}
\usepackage{mathtools}
\usepackage{verbatim}
\usepackage{enumerate}
\usepackage{graphicx}
\graphicspath{figs/}
\usepackage[colorlinks,linkcolor=red,anchorcolor=red,urlcolor=red,citecolor=red]{hyperref}
\usepackage[capitalise]{cleveref}
\usepackage{bbm}
\usepackage{booktabs}
\usepackage{bookmark}

\usepackage{subfig}
\usepackage{caption}

\usepackage{color}

\definecolor{dred}{rgb}{.8,0.2,.2}
\definecolor{ddred}{rgb}{.8,0.5,.5}
\definecolor{dblue}{rgb}{.2,0.2,.8}
\definecolor{dgreen}{rgb}{.2,0.5,.2}

\newcommand{\bra}[1]{\mbox{$\langle #1|$}}
\newcommand{\ket}[1]{\ensuremath{|#1\rangle}}

\newcommand{\be}{\begin{equation}}
\newcommand{\ee}{\end{equation}}
\newcommand{\bea}{\begin{eqnarray}}
\newcommand{\eea}{\end{eqnarray}}

\begin{document}

\title{Learning Quantum Hamiltonians from Single-qubit Measurements}

\author{Liangyu Che}
\affiliation{Shenzhen Institute for Quantum Science and Engineering and Department of Physics, Southern University of Science and Technology, Shenzhen 518055, China}
\affiliation{Guangdong Provincial Key Laboratory of Quantum Science and Engineering, Southern University of Science and Technology, Shenzhen, 518055, China}

\author{Chao Wei}
\affiliation{Shenzhen Institute for Quantum Science and Engineering and Department of Physics, Southern University of Science and Technology, Shenzhen 518055, China}
\affiliation{Guangdong Provincial Key Laboratory of Quantum Science and Engineering, Southern University of Science and Technology, Shenzhen, 518055, China}

\author{Yulei Huang}
\affiliation{Shenzhen Institute for Quantum Science and Engineering and Department of Physics, Southern University of Science and Technology, Shenzhen 518055, China}
\affiliation{Guangdong Provincial Key Laboratory of Quantum Science and Engineering, Southern University of Science and Technology, Shenzhen, 518055, China}

\author{Dafa Zhao}
\affiliation{State Key Laboratory of Low-Dimensional Quantum Physics and Department of Physics, Tsinghua University, Beijing 100084, China}

\author{Shunzhong Xue}
\affiliation{State Key Laboratory of Low-Dimensional Quantum Physics and Department of Physics, Tsinghua University, Beijing 100084, China}

\author{Xinfang Nie}
\affiliation{Shenzhen Institute for Quantum Science and Engineering and Department of Physics, Southern University of Science and Technology, Shenzhen 518055, China}
\affiliation{Guangdong Provincial Key Laboratory of Quantum Science and Engineering, Southern University of Science and Technology, Shenzhen, 518055, China}

\author{Jun Li}
\email{lij3@sustech.edu.cn}
\affiliation{Shenzhen Institute for Quantum Science and Engineering and Department of Physics, Southern University of Science and Technology, Shenzhen 518055, China}
\affiliation{Guangdong Provincial Key Laboratory of Quantum Science and Engineering, Southern University of Science and Technology, Shenzhen, 518055, China}

\author{Dawei Lu}
\email{ludw@sustech.edu.cn}
\affiliation{Shenzhen Institute for Quantum Science and Engineering and Department of Physics, Southern University of Science and Technology, Shenzhen 518055, China}
\affiliation{Guangdong Provincial Key Laboratory of Quantum Science and Engineering, Southern University of Science and Technology, Shenzhen, 518055, China}

\author{Tao Xin}
\email{xint@sustech.edu.cn}
\affiliation{Shenzhen Institute for Quantum Science and Engineering and Department of Physics, Southern University of Science and Technology, Shenzhen 518055, China}
\affiliation{Guangdong Provincial Key Laboratory of Quantum Science and Engineering, Southern University of Science and Technology, Shenzhen, 518055, China}

\begin{abstract}

It is natural to measure the observables from the Hamiltonian-based quantum dynamics, and its inverse process that Hamiltonians are estimated from the measured data also is a vital topic. In this work, we propose a recurrent neural network to learn the parameters of the target Hamiltonians from the temporal records of single-qubit measurements.
The method does not require the assumption of ground states and only measures single-qubit observables. It is applicable on both time-independent and time-dependent Hamiltonians and can simultaneously capture the magnitude and sign of Hamiltonian parameters. Taking quantum Ising Hamiltonians with the nearest-neighbor interactions as examples, we trained our recurrent neural networks to learn the Hamiltonian parameters with high accuracy, including the magnetic fields and coupling values. The numerical  study also shows that our method has good robustness against the measurement noise and decoherence effect. Therefore, it has widespread applications in estimating the parameters of quantum devices and characterizing the Hamiltonian-based quantum dynamics.
\end{abstract}

\maketitle

\section{Introduction}

Developing the methods for estimating Hamiltonians has two important motivations in quantum information processing. First, Hamiltonians fully govern the dynamics of quantum systems. Hence, whether the Hamiltonians can be precisely estimated determines whether the control operations are highly accurate on these quantum devices. For instance, quantum circuits are generally realized through the control pulse techniques \cite{schafer2018fast}, which are beforehand designed and optimized according to the parameters of the system Hamiltonians. Second, as a branch of quantum process tomography \cite{PhysRevA.77.032322}, estimating Hamiltonians provides an alternative approach to estimate the fidelities of the performed quantum simulations. Therefore, estimating Hamiltonians is a central problem in the related quantum fields, such as quantum platforms \cite{mermin2007quantum}, quantum control \cite{dong2010quantum, helsen2020general}, and quantum simulations \cite{RevModPhys.86.153, xin2020quantum}.

So far, various methodologies have been studied for this purpose. In principle, Hamiltonians can be estimated by quantum state and process tomography by considering the Hamiltonians are the generators of the dynamical processes \cite{PhysRevA.77.032322,xin2020improved,xin2017quantum}. However, this approach requires exponential physical resources, although many-body Hamiltonians have the polynomial number of unknown parameters because of the physical constraints. Previously, some methods using Fourier transform or fitting on the temporal records of measurement of some observables also are proposed to estimate Hamiltonians with few qubits \cite{PhysRevLett.102.187203,PhysRevA.71.062312,PhysRevA.73.052317}. Zhang and Sarvoar \cite{PhysRevLett.113.080401,PhysRevA.91.052121} proposed an approach for estimating Hamiltonian based on the limited measurements by the eigensystem realization algorithm (ERA). This method was experimentally demonstrated on nuclear magnetic resonance quantum processor \cite{hou2017experimental}.  Akira Sone {\it{et al}} further studied the identifiability problem of Hamiltonians and the necessary experimental resources in ERA method and they show that more observables are necessary and the required experimental measurements have exponential scaling with the size of systems for complicated Hamiltonians \cite{PhysRevA.95.022335,sone2017exact}. 
Many-body local Hamiltonians can be uniquely estimated by a single eigenstate of Hamiltonians, which also inspires the subsequent research \cite{qi2019determining,PhysRevB.100.134201,PhysRevX.8.021026, PhysRevLett.122.020504, xin2019local}. 
Recently, a quantum quench method also was proposed to reconstruct a generic many-body local Hamiltonian \cite{PhysRevLett.124.160502}, which uses pairs of generic initial and final states connected by the time evolution of Hamiltonians.

Machine learning has obtained great successes in solving the problems in quantum physics \cite{rem2019identifying,van2017learning,huembeli2018identifying,rodriguez2019identifying,lian2019machine,lu2018separability,harney2020entanglement,torlai2018neural,ahmed2020quantum,magesan2015machine,khanahmadi2020time,cimini2020neural}, such as the identification of quantum phase transitions \cite{rem2019identifying,van2017learning,huembeli2018identifying}, the classification of quantum topological phases and quantum entanglement \cite{rodriguez2019identifying,lian2019machine,lu2018separability,harney2020entanglement}, quantum state measurement and tomography \cite{magesan2015machine,torlai2018neural,ahmed2020quantum}. Recently, machine learning also brings developments in estimating the Hamiltonians. Ref. \cite{xin2019local} presents the deep neural network to recover 2-local Hamiltonians from merely 2-local measurements of ground states. Ref. \cite{cnnran}  proposes that convolutional neural networks can also be used to predict the physical parameters of Hamiltonians from the ground states. However, these methods usually require the assumption of the ground states.  

In this work, we propose a machine learning method, Recurrent Neural Network (RNN), to estimate the parameters of Hamiltonians from single-qubit Pauli measurements on each qubit.  In our method, the initial state does not require the ground states of target Hamiltonians and only single-qubit Pauli observables are measured at a discrete-time forming the temporal records of single-qubit measurements which are fed into RNN. The intuition of this method is that if the Hamiltonians are identifiable under the temporal records of single-qubit measurements, then there exists the underlying rule from single-qubit measurements to the target Hamiltonians, which can be learned directly from single-qubit measurements via data-driven machine learning, although this rule may have complicated or even unknown functional forms. Our paper is organized as follows. In Sec. \ref{sec2}, we first describe our framework for estimating Hamiltonians via RNN and then test our methods for different types of time-independent and time-dependent Hamiltonians with up to 7 qubits. The robustness against the measurement noise and decoherence is successively studied. In Sec. \ref{sec3}, we in detail discuss the required measurement resources in the practical applications, followed by our conclusions and outlooks. The detailed techniques in our method are placed in Sec. \ref{sec4}.

\section{Results}\label{sec2}
\subsection{Learning Hamiltonians via the RNN}
We firstly describe the dynamics of single-qubit observables under the target Hamiltonians. Here, we consider that a quantum system with $N$ qubits starts from an initial state $\ket{\Psi_0}$ and undergoes a  dynamical process governed by the unknown Hamiltonian $\mathcal{H}$.  $\mathcal{H}$ is parameterized as
\begin{equation}
\begin{aligned}
\mathcal{H}=\sum_{m=1}^M a_m B_m,
\end{aligned}
\end{equation}
where $B_m$ is the tensor product of Pauli matrices $I, \sigma_x, \sigma_y,$ and $\sigma_z$, and $a_m$ is  the parameter of Hamiltonians. For single-qubit Pauli operator $P\in S_P=\{\sigma_k^{(i)}|k=x, y, z, 1\leqslant i\leqslant N\}$, its expectation value is $\overline{P}(t)=\bra{\Psi_0}P(t)\ket{\Psi_0}$ with $P(t)=e^{i\mathcal{H}t}Pe^{-i\mathcal{H}t}$. Here, $P(t)=\sum_{n=0}\frac{i^nt^n}{n!}P_n$. $P_0=P$ and $P_n=\sum_{m=1}^M a_m [B_m, P_{n-1}]$. Hence, if the parameter $a_m$ participates in the dynamics of single-qubit observables, it is possible to learn the Hamiltonian parameters from the temporal records of their expectation values.  In this work, we consider the identifiable Hamiltonians under single-qubit measurements and the initial state $\ket{\Psi_0}$. The most common of Hamiltonians belong to this category. Next, we describe our machine learning method for estimating the Hamiltonians from single-qubit Pauli measurements.

\begin{figure}[h]
\centering
\includegraphics[width=1\linewidth]{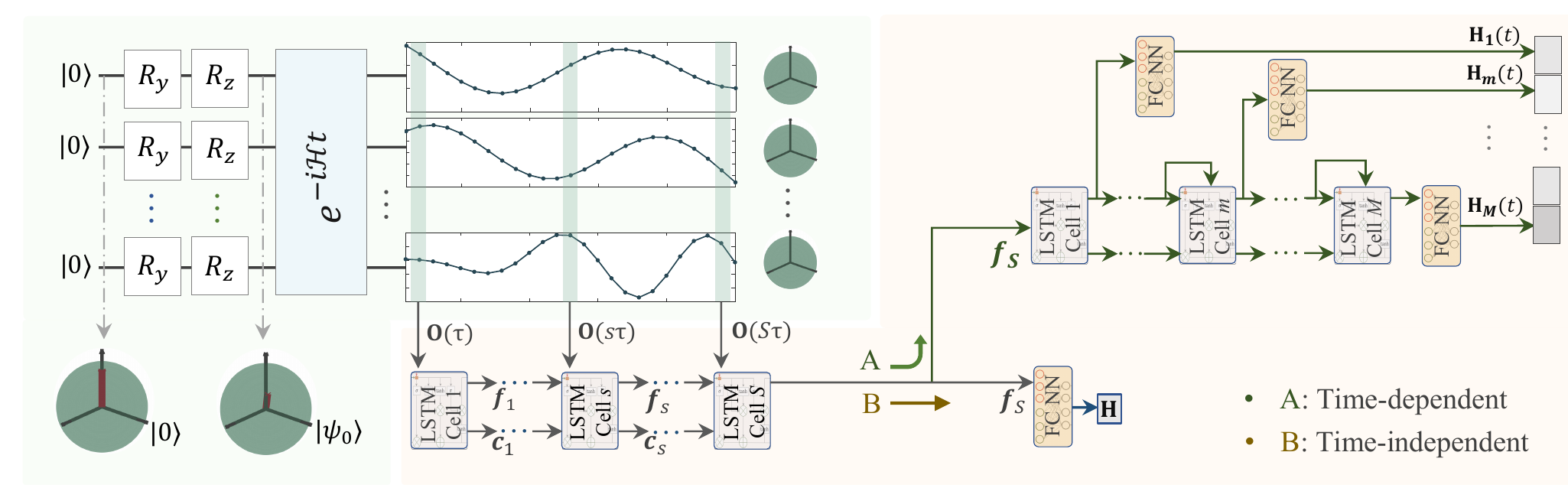}
\caption{\textbf{Circuit diagram of our neural networks on learning the parameters of Hamiltonians from the temporal records of single-qubit measurements.} We first perform the dynamical evolution $e^{-i\mathcal{H}t}$ from the initial state $\ket{\psi_0}$ (bottom left Bloch). At each moment $s\tau$, we measure the expectation values of single-qubit Pauli operators (middle Bloch), and they are collected as a vector $\textbf{O}(s\tau)$ fed into $s$-th LSTM cell. Lastly, the combination of FC and LSTM for time-independent parameters in Hamiltonians (path A) or an FC neural network for time-independent parameters in Hamiltonians (path B) follows LSTM cells.}
\label{pro}
\end{figure}

As illustrated in Fig. \ref{pro}, a $N$-qubit system starts from the initial state $\ket{\Psi_0}=\prod_{i=1}^N \otimes \ket{\psi^i_0}$. Here, $\ket{\psi^i_0}=R_z(\pi/4)R_y(\pi/4)\ket{0}$ can be prepared from the state $\ket{0}$ using rotation operations $R_z(\pi/4)$ and $R_y(\pi/4)$. The purpose for choosing such an initial state ensures that the dynamics of single-qubit observables have nontrivial initial values. During the dynamical evolution $e^{-i\mathcal{H}t}$, the expectation values of single-qubit operators $\sigma_x^{(i)}$, $\sigma_y^{(i)}$, and $\sigma_z^{(i)}$ are measured at a discrete-time by time interval $\tau$. Total sample points is denoted by $S$ and then total sample time is $T=S\tau$. The temporal records of single-qubit measurements are collected as a vector,
\begin{equation}
\begin{aligned}
\textbf{I}=\{O^{(i)}_k(s\tau) | O^{(i)}_k(s\tau)=\text{Tr}(\rho(s\tau)\cdot \sigma^{(i)}_k), 1\leqslant s \leqslant S, k = x, y, z,1\leqslant i \leqslant N\}.
\end{aligned}
\end{equation}
$\rho(s\tau)$ is the density matrix of the system at the moment $s\tau$. The parameters of Hamiltonians are collected as a vector $\textbf{H}=\{a_m | 1\leqslant m \leqslant M\}$. Then we can train a neural network framework consisting of Recurrent and full connected (FC) neural networks with generated training data $\{\textbf{I}, \textbf{H}\}$. After the training, we can predict the unknown Hamiltonian parameters $\textbf{H}$ from single-qubit measurements $\textbf{I}$. As shown in Fig. \ref{pro}, in our NN framework, we use Long Short-Term Memory (LSTM) network which is a type of RNN \cite{hochreiter1997long}. Compared with traditional feed-forward neural networks, LSTM can learn the correlation in time sequences, which has been widely applied on handwriting recognition and speech recognition in the classical field \cite{sak2014long}, and quantum control and quantum process tomography in the quantum field \cite{banchi2018modelling,PhysRevX.10.011006}. LSTM is appropriate to estimate the Hamiltonians from the temporal records.  In this training, we define the input and output layers, objective function, and similarity function as follows.

 (i) The input and output layers. $\textbf{I}$ and $\textbf{H}$ are respectively used as the input and output layers of our NN framework. At the moment $s\tau$, the expectation values of single-qubit measurements are collected as a vector
 \begin{equation}
\begin{aligned}
\textbf{O}(s\tau)=\{O^{(i)}_k(s\tau)| k = x, y, z,~1\leqslant i \leqslant N\}.
\end{aligned}
\end{equation}
 It is firstly fed into the $s$-th LSTM cell. Lastly, an FC neural network is applied before exporting the prediction $\textbf{H}$. Hence, the number of required LSTM cells equals the number of sampling points $S$. \\
 (ii) The objective function. Our neural network is trained by minimizing the distance between the predicted outcome $\textbf{H}^{\text{pred}}$ and the true outcome $\textbf{H}^{\text{true}}$. Here, we use Mean Square Error (MSE) between $\textbf{H}^{\text{pred}}$ and $\textbf{H}^{\text{true}}$ as the objective function. It is
 \begin{equation}
\begin{aligned}
L=\frac{1}{M}\sum_{m=1}^M(\textbf{H}^{\text{true}}_m-\textbf{H}^{\text{pred}}_m)^2.
\label{loss}
\end{aligned}
\end{equation}
 This definition can learn the magnitude and sign of the parameters, because $L$ decreases to 0 only when $\textbf{H}^{\text{true}}_m$ and $\textbf{H}^{\text{pred}}_m$ are absolutely the same. To minimize the objective function in this work, we use Adam optimization algorithm, one of the state-of-the-art gradient descent algorithms, to train the hidden parameters of the network. \\
 (iii) The similarity function. To estimate the performance of our trained NN, we need to compute the similarity between the predicted and the real outcomes for the test data. Here, we use the definition of the cosine proximity function between two vectors. It is
  \begin{equation}
\begin{aligned}
F(\textbf{H}^{\text{pred}}, \textbf{H})=\frac{(\textbf{H}^{\text{pred}}\cdot\textbf{H})}{(||\textbf{H}^{\text{pred}}|| \cdot ||\textbf{H}||)}.
\end{aligned}
\end{equation}
Here, $F \in [-1, 1]$. As shown in Fig. \ref{pro}, the structures for exporting $\textbf{H}$ are different for time-dependent and time-independent Hamiltonians. For time-dependent parameters in Hamiltonians, $\emph{f}_S$ is imported to a composite neural network including LSTM and FC neural networks (path A). Repetitive LSTM cells decode the vector $\emph{f}_S$ and FC neural networks project the output of each cell to a series of time-dependent Hamiltonian parameters. For time-independent parameters in Hamiltonians, an FC neural network directly follows the LSTM cells. Here, the FC neural networks do not have hidden layers. More details about the structure of LSTM can be found in Methods \ref{sec4}. Next, we will train neural networks to learn the parameters of different types of Hamiltonians.

\subsection{Applications}\label{DFE}

{\it{Ising Hamiltonian 1}}-. As a demonstration of applications, we first train an RNN framework for estimating the parameters of 7-qubit Hamiltonians with the nearest-neighbor XY interactions placed in a static magnetic field around $z$ axis as follows,
\begin{equation}
\begin{aligned}
\mathcal{H}^7_{\text{XYZ}}=\sum^7_{i=1}a^{(i)}_z \sigma_z^{(i)}+\sum_{j=1}^6 J^{(j)}(\sigma_x^{(j)}\sigma_x^{(j+1)}+\sigma_y^{(j)}\sigma_y^{(j+1)}).
\label{xyz}
\end{aligned}
\end{equation}
$a^{(i)}_z$ and $J^{(j)}$ are the parameter of magnetic field on $j$-th qubit and the coupling value between the nearest-neighbor qubits, respectively. Suppose that $a^{(i)}_z\in [-J_0, J_0]$ and $J^{(j)}\in [-J_0, J_0]$. $J_0$ is a global factor which is set to 1 in our training. The system evolves under the Hamiltonian $\mathcal{H}^7_{\text{XYZ}}$ starting from the initial state $\prod_{i=1}^7 \otimes \ket{\psi_0}$, and the expectation values of single-qubit observables $\sigma_x^{(i)}$, $\sigma_y^{(i)}$, and $\sigma_z^{(i)}$ are measured at a discrete-time separated by $\tau=0.02\pi/J_0$ with $S=25$ sampling points. The reason for choosing such a time interval can be found in Sec. \ref{sec3}. 

\begin{figure*}[htp]
\centering
\includegraphics[width=1\linewidth]{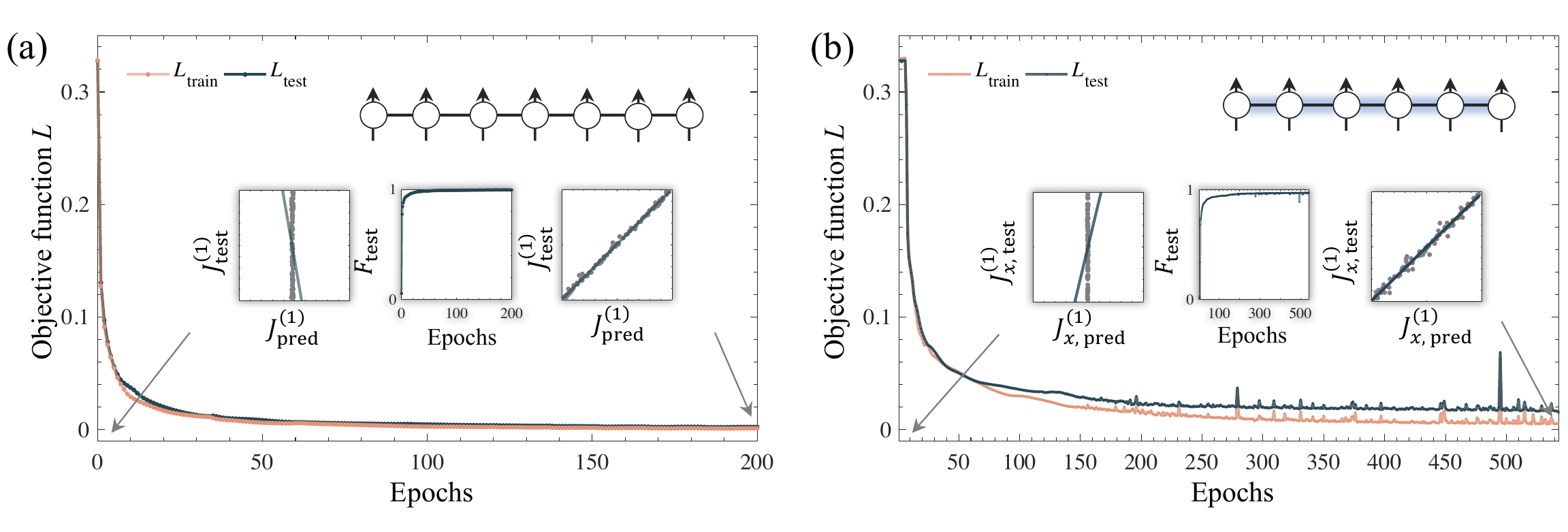}
\caption{\textbf{Trained results for 7-qubit Ising Hamiltonian 1 (a) and 6-qubit Ising Hamiltonian 2 (b).} The top right corners of panels respectively present their qubit configurations. The orange and cyan lines show the objective functions $L_{\text{train}}$ and $L_{\text{test}}$ as a function of epochs. The similarity $F_{\text{test}}$ between the predicted $\textbf{H}^{\text{pred}}$ and the true $\textbf{H}^{\text{true}}$ in the test data is also presented with the increase of epochs (middle subfigures). At the beginning and end of the training, we randomly choose 100 samples and plot the comparison between the predicted and actual values for the parameters $J^{(1)}$ (left and right subfigures).}
\label{7qubit}
\end{figure*}

We collect the Hamiltonian parameters as a vector $\textbf{H}=\{a^{(i)}_z, J^{(j)} | 1\leqslant i \leqslant 7, 1\leqslant j \leqslant 6\}$ and the measured values as a vector $\textbf{I}=\{O^{(i)}_k(s\tau) | O^{(i)}_k(s\tau)=\text{Tr}(\rho(s\tau)\cdot \sigma^{(i)}_k), 1\leqslant s \leqslant 25, k = x, y, z, \text{and} ~1\leqslant i \leqslant 7\}$, and then we randomly generate 100,000 training data $\{\textbf{I}, \textbf{H}\}$ fed into the neural networks. The test data consists of 5,000 pairs of Hamiltonians $\textbf{H}$ and the corresponding single-qubit measurements $\textbf{I}$. Our RNN is trained by minimizing the distance between the actual and predicted outcomes in Eq. (\ref{loss}). After finishing the training of RNN on the training data, our RNN has the ability to estimate the unknown parameters of 7-qubit Hamiltonians $\mathcal{H}^7_{\text{XYZ}}$ from single-qubit measurements with high accuracy. We compute the similarity $F_{\text{test}}$ between the actual parameters $\textbf{H}^{\text{test}}$ and the predicted outcome $\textbf{H}^{\text{pred}}$ for 5,000 test data. The averaged similarity on the whole test data is over 0.99 and $F_{\text{test}}$ as a function of epochs is also presented in Fig. \ref{7qubit}(a). Figure \ref{7qubit}(a) also gives the comparison between the actual value $J^{(1)}_{\text{test}}$ and the prediction $J^{(1)}_{\text{pred}}$ for 100 randomly test data at the beginning and end.

{\it{Ising Hamiltonian 2}}-. Besides, our RNN framework can also be applied to more general Hamiltonian models. Here, we use our RNN to learn the parameters of 6-qubit Ising Hamiltonians with the nearest-neighbor interactions in three directions. The Hamiltonian of this 6-qubit system can be written as,
\begin{equation}
\begin{aligned}
\mathcal{H}=\sum_{i=1}^6 a^{(i)}_z\sigma_z^{(i)}+\sum_{i=1}^{5}(J^{(i)}_x\sigma_x^{(i)}\sigma_x^{(i+1)}+J^{(i)}_y\sigma_y^{(i)}\sigma_y^{(i+1)}+J^{(i)}_z\sigma_z^{(i)}\sigma_z^{(i+1)})
\end{aligned}
\end{equation}
Similarly, single-qubit observables $\sigma_x^{(i)}$, $\sigma_y^{(i)}$, and $\sigma_z^{(i)}$ also are measured at a discrete-time separated by $\tau=0.02\pi/J_0$ and the number of sampling points is $S=75$. We randomly generate 200,000 pairs of such Hamiltonians and the corresponding single-qubit measurements as the training data. After learning on these training data, RNN can predict the outcome of the test data. For 5,000 randomly generated test data, the average accuracy of the predictions is around 0.98.  More details about the results can be found in Fig. \ref{7qubit}(b).

\begin{figure}[htp]
\centering
\includegraphics[width=0.95\linewidth]{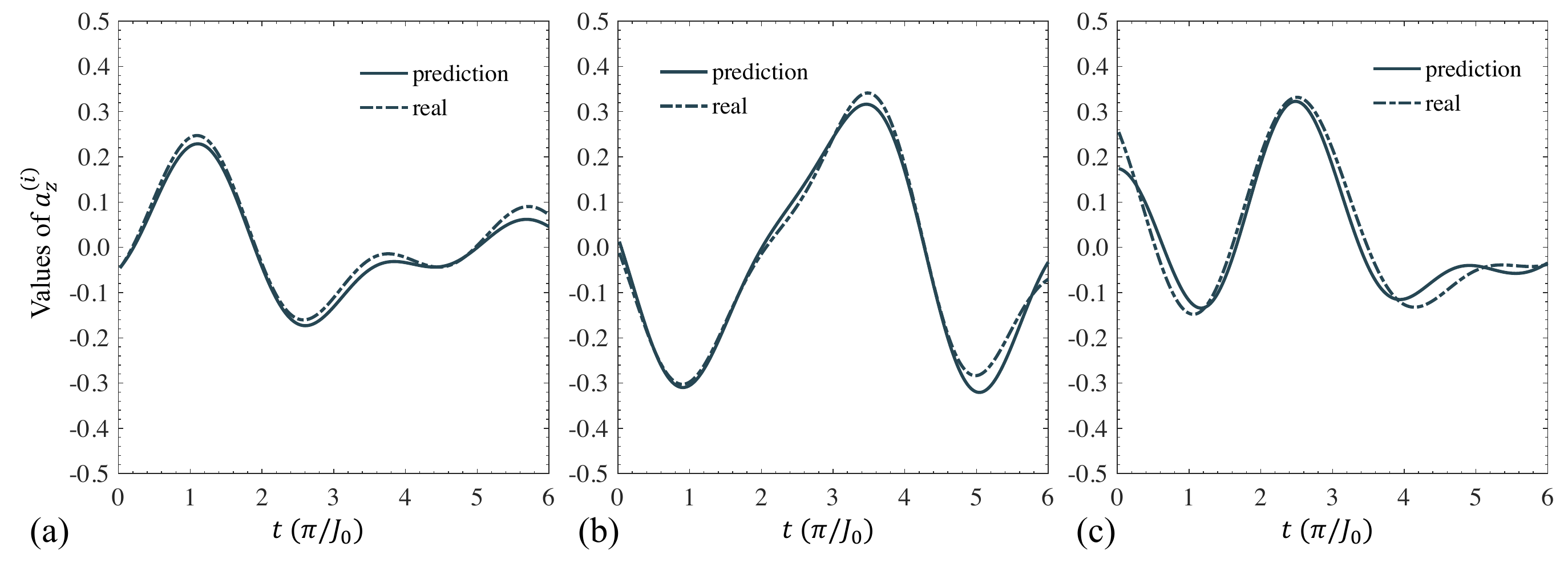}
\caption{\textbf{The temporal curves of the actual parameters (dotted lines) and the values learned by RNN (solid lines) for time-dependent parameters $a^{(i)}_z(t)$.} The predictions of time-independent parameters are $J^{(1)}_{\text{pred}}=0.0464$  ($J^{(1)}_{\text{true}}=0.0326$) and $J^{(2)}_{\text{pred}}=-0.0345$  ($J^{(2)}_{\text{true}}=-0.0181$). }
\label{plottime}
\end{figure}

{\it{Time-dependent Hamiltonians}}-. Most of the existing methods are designed for the time-independent Hamiltonians and they are not directly applicable to time-dependent Hamiltonians. Our proposed RNN method presented in the above can also be  used to estimate the parameters of time-dependent Hamiltonians. As a numerical demonstration, we consider a 3-qubit system with the nearest-neighbor XY interactions placed in a time-dependent  magnetic field around $z$ axis. The used neural network is presented in Fig. \ref{pro}. The corresponding Hamiltonian is,
\begin{equation}
\begin{aligned}
\mathcal{H}^3_{\text{XYZ}}(t)=\sum^3_{i=1}a^{(i)}_z(t) \sigma_z^{(i)}+\sum_{j=1}^2 J^{(j)}(\sigma_x^{(j)}\sigma_x^{(j+1)}+\sigma_y^{(j)}\sigma_y^{(j+1)}).
\end{aligned}
\end{equation}
We assume that $a^{(i)}_z(t)$ is a random combination of $W$ Fourier series, $a^{(i)}_z(t)=\frac{1}{W}\sum^W_{w=1}F_w\text{cos}(\nu_w t+\phi_w)$ and $J_i\in[-J_0, J_0]$ is static in time . $F_w\in[-J_0, J_0]$, $\nu_w\in[-J_0, J_0]$, and $\phi_w\in[0, 2\pi]$ are the amplitude, frequency, and phase of $w$-th series, respectively. In this case, we set $W=10$. The parameters of $\mathcal{H}^3_{\text{XYZ}}(t)$ are collected as a vector $\textbf{H}=\{a^{(i)}_z(s\tau), J^{(j)}|1\leqslant s \leqslant 300, 1\leqslant i \leqslant 3, 1\leqslant j \leqslant 2\}$. The expectation values of single-qubit observables also are measured at a discrete time separated by $\tau=0.02\pi/J_0$, and they are collected as a vector $\textbf{I}=\{O^{(i)}_k(s\tau) | O^{(i)}_k(s\tau)=\text{Tr}(\rho(s\tau)\cdot \sigma^{(i)}_k), 1\leqslant s \leqslant 300, k = x, y, z, \text{and} ~1\leqslant i \leqslant 3\}$. Our training data also consists of 100,000 randomly generated pairs of Hamiltonians $\textbf{H}$ and the corresponding single-qubit measurements $\textbf{I}$. After training RNN to convergence on these training data, it can be used to learn the temporal behavior of $a^{(i)}_z(t)$ from only the measurements $\textbf{I}$. Figure \ref{plottime} presents the temporal behavior of the predicted values (solid lines) and its comparison with the actual values (dotted lines) for time-dependent parameters $a^{(i)}_z(t)$. It shows that a good agreement between the predicted and real results has been achieved.

\subsection{Robustness against the noise}\label{DFE}

The temporal records of single-qubit measurements inevitably are influenced by the statistical and environmental noises, and may these noises deviate the predicted values of RNN from the ideal ones. Here, we further study the robustness of our RNN framework in learning the unknown Hamiltonians under the Gaussian noise and decoherence effect. Following simulations are performed for a 3-qubit system with Ising Hamiltonian  $\mathcal{H}^3_{\text{XYZ}}$. $\mathcal{H}^3_{\text{XYZ}}$ is
\begin{equation}
\begin{aligned}
\mathcal{H}^3_{\text{XYZ}}=\sum^3_{i=1}a^{(i)}_z \sigma_z^{(i)}+\sum_{j=1}^2 J^{(j)}(\sigma_x^{(j)}\sigma_x^{(j+1)}+\sigma_y^{(j)}\sigma_y^{(j+1)}).
\end{aligned}
\label{3q}
\end{equation}
The unknown parameters in $\mathcal{H}^3_{\text{XYZ}}$ form a vector $\textbf{H}=[a^{(1)}_z, a^{(2)}_z, a^{(3)}_z, J^{(1)}, J^{(2)}]^T$ as the output of RNN. The expectation values of single-qubit observables $\sigma_x^{(i)}, \sigma_y^{(i)}$, and $ \sigma_z^{(i)}$ are measured at a discrete time  separated by $\tau=0.02\pi/J_0$, and they are collected as the input data $\textbf{I}$ of RNN.

\begin{figure}[htp]
\centering
\includegraphics[width=0.85\linewidth]{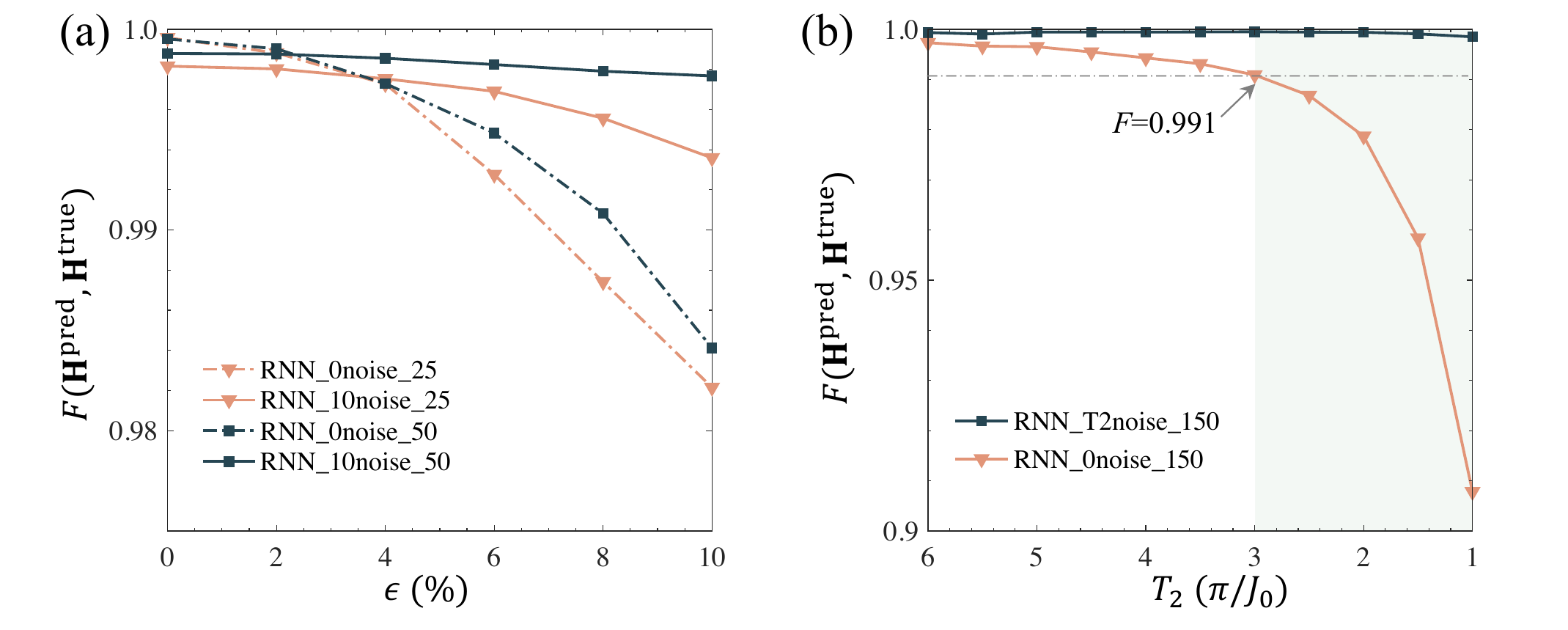}
\caption{\textbf{The numerical simulated results for the robustness.} (a) The predicted accuracy of trained RNN models ($\texttt{\text{RNN}\_\text{0noise}\_\text{25}}$, $\texttt{\text{RNN}\_\text{10noise}\_\text{50}}$, $\texttt{\text{RNN}\_\text{0noise}\_\text{25}}$, and $\texttt{\text{RNN}\_\text{10noise}\_\text{50}}$) under the influence of Gaussian noise. (b) The predicted accuracy of trained RNN models ($\texttt{\text{RNN}\_\text{T2noise}\_\text{150}}$ and $\texttt{\text{RNN}\_\text{0noise}\_\text{150}}$) under the influence of decoherence effect. The cyan shadow is that the sampling time is longer than coherence time. }
\label{plotnoise}
\end{figure}

{\it{Robustness against the Gaussian noise}}-. First, we train RNN frameworks by feeding 100,000 noiseless training data $\{\textbf{I}, \textbf{H}\}$ with the sampling points $S=25$ and $S=50$, respectively.  Two trained RNN models $\texttt{\text{RNN}\_\text{0noise}\_\text{25}}$ and $\texttt{\text{RNN}\_\text{0noise}\_\text{50}}$ are obtained. Then we predict the Hamiltonian parameters by feeding noisy test data into these two RNN models.  These noisy data is artificially generated by adding the Gaussian noise in the data $\textbf{I}$, i.e., $\textbf{I}'=\textbf{I}+\mathcal{N}(0, \epsilon)$. Here, $\mathcal{N}(0, \epsilon)$ is a Gaussian distribution with the mean of 0 and the standard deviation of $\epsilon$. We change $\epsilon$ from 2\% to 10\% with the step 2\% and create 5,000 noisy test data for each $\epsilon$. Figure \ref{plotnoise}(a) presents the average similarities between the predicted parameters $\textbf{H}^{\text{pred}}$ and the true parameters $\textbf{H}^{\text{true}}$as a function of $\epsilon$. $\texttt{\text{RNN}\_\text{0noise}\_\text{50}}$ has a better performance than $\texttt{\text{RNN}\_\text{0noise}\_\text{25}}$, but both their predicted accuracy decrease with the increasing of $\epsilon$. When $\epsilon=0.1$, the accuracy of $\texttt{\text{RNN}\_\text{0noise}\_\text{25}}$ decreases to 0.98. To further improve the robustness of our RNN frameworks under the noise, we adopt the following approach.

Second, we change to train RNN frameworks by feeding 100,000 noisy training data. The training data is perturbed under a Gaussian noise with a standard deviation of $\epsilon=0.1$.  Two RNN models $\texttt{\text{RNN}\_\text{10noise}\_\text{25}}$ and $\texttt{\text{RNN}\_\text{10noise}\_\text{50}}$ are trained to convergence. Similarly, we use these models to test the noisy data. $\epsilon$ also is changed from 2\% to 10\% with the step 2\% and 5,000 noisy test data for each $\epsilon$ is created.  The average values of  predicted accuracy as a function of $\epsilon$ also are  presented in Fig. \ref{plotnoise}(a). It shows that it has good performance with the similarity of over 0.99 and the predicted accuracy improves to 0.995 from the previous 0.98 when $\epsilon=0.1$. The above simulations show that training RNN frameworks with the noisy data will greatly enhance the predicted accuracy and the more sample points will bring better robustness against the noise. From the simulation, it can be roughly concluded that learning Hamiltonians via RNN has robust performance under the Gaussian noise.

{\it{Robustness against the decoherence}}-. The total time for measuring the temporal records may reach or even exceed the coherence time of the experimental devices. Hence, the collected temporal records contain the decoherence effect, leading to a decrease in the predicted accuracy. For this purpose, we also numerically study the performance of our RNN frameworks under the decoherence effect. The temporal records with decoherence effect are created according to the Kraus representation of decoherence dynamics. The evolution of Hamiltonians is divided into slices with the duration of each slice being $\delta\tau$. Supposing that the density matrix is $\rho(t)$ at the moment $t$, then density matrix at $t+\delta\tau$ is
\begin{equation}
\begin{aligned}
\rho(t+\delta\tau)=\sum_{i=1}^3\sum_{j=0}^1 E^i_j e^{-i\mathcal{H}\delta\tau} \rho(t)e^{i\mathcal{H}\delta\tau}  E^{i\dag}_j.
\end{aligned}
\end{equation}
Here, $ E^i_j $ is the kraus operator of the $i$-th qubit with,
\begin{equation}
\begin{aligned}
E^i_0=\sqrt{\lambda_i}I_2, E^i_1= \sqrt{1-\lambda_i}\sigma_z^i
 \end{aligned}
\end{equation}
$\lambda_i$ is a parameter with $\lambda_i=(1+e^{-\delta \tau/T_2^i})/2$.  $T_2^i$ is the decoherence time of $i$-th qubit. We change $T_2^i$ from $1\pi/J_0$ to $6\pi/J_0$ with the segment $2\pi/J_0$. For each $T_2^i$, we create 5,000 decoherence test data with the sample points of $S=150$ (Sample interval is $0.02\pi/J_0$ and corresponding sampling time is $3\pi/J_0$). 

As shown in Fig. \ref{plotnoise}(b), when we feed these test data to the model $\texttt{\text{RNN}\_\text{0noise}\_\text{150}}$ to predict the Hamiltonian parameters $\textbf{H}$, it is found that the accuracy of predicted $\textbf{H}$ rapidly falls with the decrease of coherence time. To improve the robustness against the decoherence effect, we trained a RNN framework using 100,000 decoherence train data, named by $\texttt{\text{RNN}\_\text{T2noise}\_\text{150}}$. Figure \ref{plotnoise}(b) presents that the predicted accuracy will have a significant improvement with the average value of over 0.99, when using $\texttt{\text{RNN}\_\text{T2noise}\_\text{150}}$ to process the decoherence test data.

\begin{figure}[htp]
\centering
\includegraphics[width=0.85\linewidth]{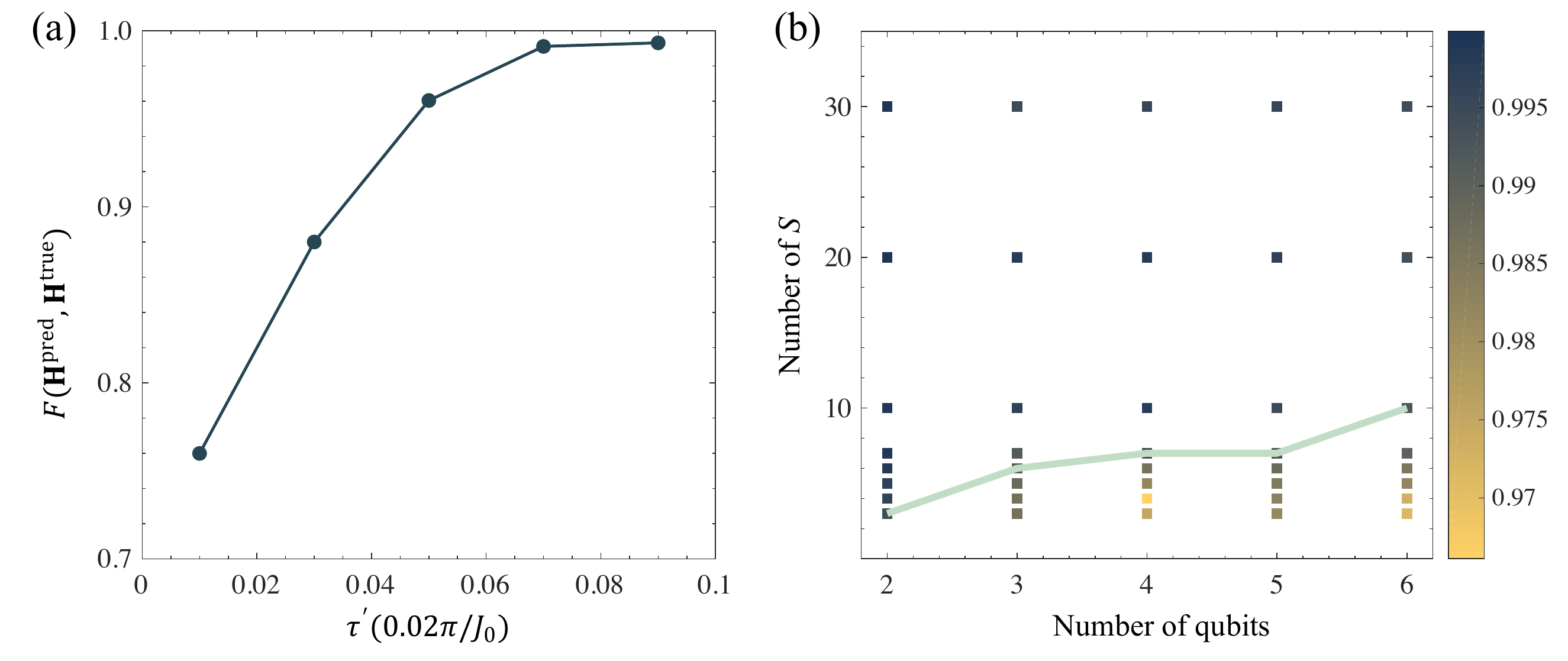}
\caption{\textbf{Numerical simulations in the discussion.} (a) The achieved accuracy with the different sampling intervals $\tau'$ and fixed sampling points $S=25$. (b) The achieved accuracy under the different number of qubits $N$ and sampling points $S$. The simulations are performed for Ising Hamiltonians in Eq. (\ref{xyz}). The cyan line is drew for the points with the accuracy of over 0.99.  }
\label{plotN}
\end{figure}

\section{Discussion and Conclusion}\label{sec3}

We briefly discuss the required measurement resources and feasibility in the practical experiments, including the sampling interval and the number of sampling points. First, single-qubit measurements are easy-to-implemented in current quantum platforms \cite{xin2020improved,keith2019single,bruzewicz2019trapped,irber2020robust,xin2018nuclear}, such as using the dispersive readout on superconducting qubits and the ensemble measurements on nuclear magnetic resonance. Single-qubit measurements also have the lower readout errors than multi-qubit measurements \cite{gambetta2007protocols,nachman2020unfolding}. Second, the sampling interval $\tau$ should be traded-off, accounting for the coherence time. On the one hand, the total sampling time may exceed the coherence time of qubits if $\tau$ is too large, leading to the decrease of the prediction accuracy. On the other hand, the temporal records of single-qubit measurements may be hard to distinguish if $\tau$ is too small, also leading to the decrease of the prediction accuracy. As shown in Fig. \ref{plotN}(a), we change the sampling interval $\tau'$ from $0.01\tau$ to $0.09\tau$ with the step $0.02\tau$ ($\tau=0.02\pi/J_0$) and fix the sampling points $S=25$. Then we train our RNN models with 100,000 training data for each $\tau'$ and test their performance with 5,000 test data. The considered Hamiltonian is described in Eq. (\ref{3q}). The result shows that the RNN model can not be trained to a high accuracy if $\tau'$ is too small. 

Third, the number of total sampling points is $3NS$, where factor 3 is the number of elements $\{\sigma_x^{(i)}, \sigma_y^{(i)}, \sigma_z^{(i)}\}$, $N$ is the number of qubits, and $S$ is the number of sample points. Here, we numerically study how $S$ increases with the size of the systems in our method. In our simulation, we consider Ising Hamiltonians in Eq. (\ref{xyz}), in which the number of qubits is changed from 2 to 6, and we train the neural networks with 100,000 randomly generated training data for given $N$ and $S$. Then, we test the average accuracy of trained neural networks using 5,000 test data. Figure \ref{plotN}(b) presents the achieved accuracy as a function of $N$ and $S$. From the simulated results, it is shown that $S$ has a gentle increasing with the size of the system for this type of Hamiltonians. It may be understood from the following aspect. As long as this Hamiltonian is identifiable under the chosen initial states and single-qubit observables, it is possible to learn the parameters of Hamiltonians from their temporal records with finite sampling points. For instance, many-body Hamiltonians have polynomial parameters. The polynomial sampling points may be enough to estimate the parameters for many-body Hamiltonians in machine learning method. 

In summary,  we conclude that a composite neural network can be trained to learn the Hamiltonians from single-qubit measurements, and numerical simulations of up to 7 qubits have demonstrated its feasibility on time-independent and time-dependent Hamiltonians. Compared with the existing methods, this neural network method does not need to prepare the eigenstates of target Hamiltonians  and it can learn all the information of Hamiltonians including the magnitude and sign of the parameters. Once the neural network is successfully trained, it can be directly used to predict the parameters of unknown Hamiltonians from
 the measured data without any post-processing. It is a `once for all' advantage. Besides, the initial states and single-qubit measurements in this method are easy-to-implemented in current quantum platforms, and the high accuracy can be achieved even under the potential experimental noises, including Gaussian noise and decoherence effect. It will bring some potential applications in performing the tasks of Hamiltonians identification in the experiments. Our method also has possible extensions in the future, such as learning the environment information around the system and simulating the dynamics of closed and open systems.

\section{Methods}\label{sec4}
\begin{figure}[h]
\centering
\includegraphics[width=0.7\linewidth]{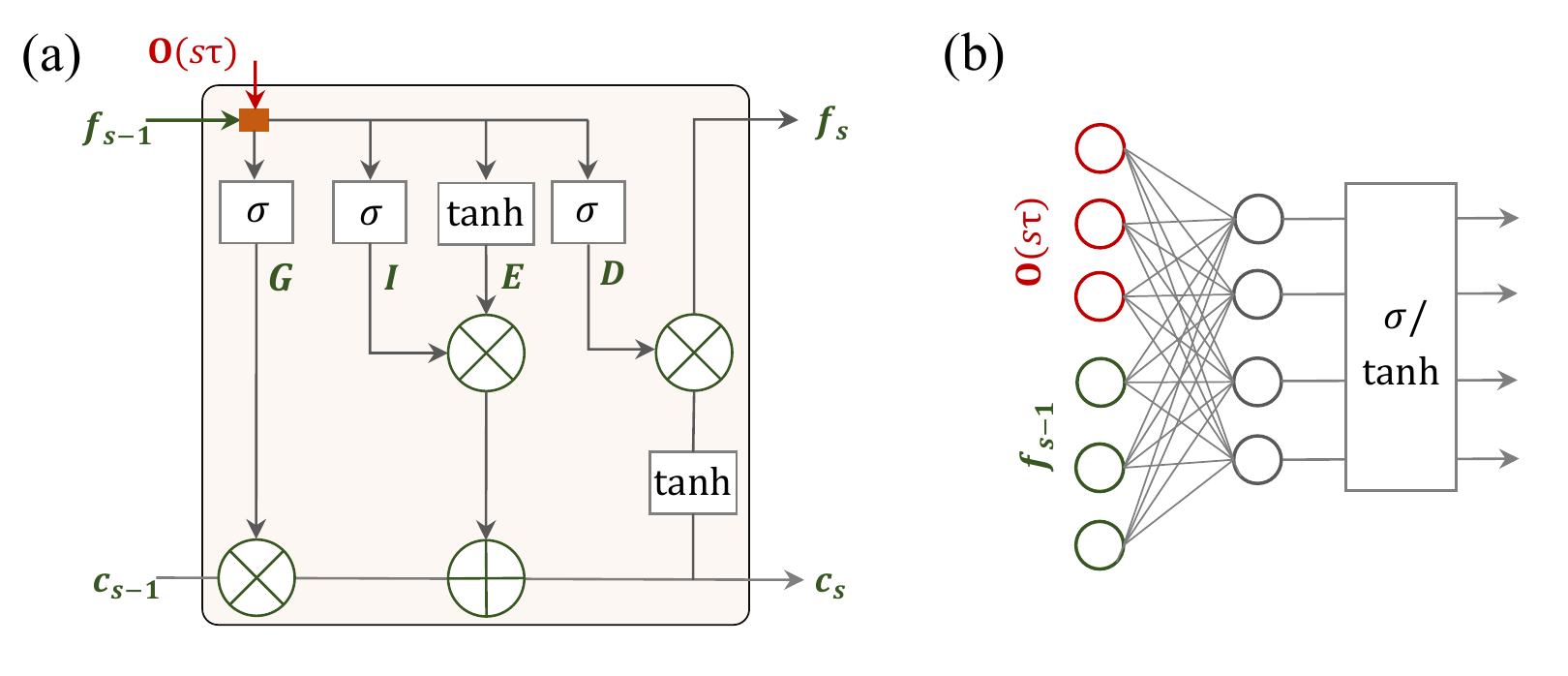}
\caption{\textbf{The schematic diagram for LSTM (a)-(b).} The right plot presents the operation combining the input $f_{s-1}$ and $O(s\tau)$ (labeled by red square) with one layer  including 256 neurons. }
\label{lstm}
\end{figure}

{\it{Structure of LSTM}}-. The LSTM is a form of the recurrent neural network designed to solve the long-term dependencies problem. An LSTM consists of a chain of repeating neural network modules called LSTM cells. As shown in Fig. \ref{lstm}(a), the $s$-th LSTM cell imports $\text{O}(s\tau)$, $f_{s-1}$, and $c_{s-1}$ and exports  $f_{s}$ and $c_{s}$ for the next LSTM cell. Here,  $\text{O}(s\tau)$ and $f_{s-1}$ are firstly combined by an FC neural network whose structure is shown in Fig. \ref{lstm}(b). In our training, this layer includes 256 neurons. Then different activation functions $\sigma$ and $\text{tanh}$ are used and finally different operations $\oplus$ and $\otimes$ are implemented before exporting $f_{s}$ and $c_{s}$. Next, we introduce the detailed operations in the LSTM cell.

As shown in Fig. \ref{lstm}, the long-term memory of LSTM is called cell state $c_s$, which stores information learned by flowing through the entire chain. To update the cell state, the cell has two layers called "forget gate" and "input gate" to remove or add information to the cell state. The cell also has the ability to output the information from cell state called "output gate". Thus, these three gates control the cell state and construct an LSTM cell. At the beginning, the cell uses forget gate $G$ to decide what past information to remove from the cell. The input of current moment $o(s)$ and the output of last moment $f_{s-1}$ go through the forget gate $G$ as follows: 
\begin{equation}
G = \sigma (W_g\cdot [f_{s-1},o(s)]^T+b_g),
\end{equation}
where $\sigma (x)= 1/(1+e^{-x})$ is the Sigmoid function. Then, it uses input gate $I$ to decide what new information to add to the cell state as follows:  $I = \sigma (W_i\cdot [f_{s-1}, o(t)]^T+b_i)$. And $o(s)$ and $f_{s-1}$ go through a tanh layer to create a candidate cell state $E$ as follows:  
\begin{equation}
E = \text{tanh}(W_e\cdot[f_{s-1}, o(s)]^T+b_e]) .
\end{equation}
The next step is to update the cell state by forget gate $G$ and input gate $I$ as follows: $c_s = G\times f_{s-1} + I \times  E $. In the end, it uses output gate to decide what information to select as output and generate the output. The equations are given as: $D = \sigma (W_d\cdot[f_{s-1}, o(s)]^T+b_d)$ and $f_s = D\times \text{tanh}(c_s)$.

\noindent {\bf Data Availability.} The experimental data and the source code that support the findings of this study can be obtained from the corresponding authors by email.

\noindent {\bf Competing Interests.} The authors declare that there are no competing interests.

\noindent {\bf Author Contributions.}  C. W. made the corresponding simulations and created the data for training the neural networks. L. C. trained the neural networks. T. X. supervised this project in this work. All the authors joined the discussions, and wrote and modified the manuscript. L. C. and C. W. contributed equally to this work.

\noindent {\bf Funding.} This work is supported by the National Key Research and Development Program of China (2019YFA0308100), National Natural Science Foundation of China (12075110, 11975117, 11905099, 11875159  and U1801661),  Guangdong Basic and Applied Basic Research Foundation (2019A1515011383), Guangdong International Collaboration Program (2020A0505100001), Guangdong Provincial Key Laboratory (2019B121203002), Science, Technology and Innovation Commission of Shenzhen Municipality (ZDSYS20170303165926217, KQTD20190929173815000, JCYJ20200109140803865, JCYJ20170412152620376 and JCYJ20180302174036418), and Pengcheng Scholars, Guangdong Innovative and Entrepreneurial Research Team Program (2019ZT08C044).
    

\end{document}